\documentstyle[prl,aps,twocolumn,multicol,epsf]{revtex} 
\tighten
\begin{document} 
\twocolumn[\hsize\textwidth\columnwidth\hsize\csname
@twocolumnfalse\endcsname 
\title{Singularity in 2+1 dimensional AdS-scalar black hole} 
\author{Lior M. Burko} 
\address{Theoretical Astrophysics,
California Institute of Technology, Pasadena, California 91125} 
\date{\today}

\maketitle

\begin{abstract}
We study the spacetime singularity in 2+1 dimensional AdS-scalar black hole
with circular symmetry using a quasi-homogeneous model. We show that this is
a spacelike, scalar curvature, deformationally strong singularity. 
\newline
\newline
PACS number(s): 04.20.Dw, 04.60.Kz, 04.70.Bw
\end{abstract}

\vspace{3ex}
]

Recently there has been growing interest in asymptotically anti de-Sitter
(AdS) black holes, motivated by the discovery of the BTZ black hole
\cite{btz} in General Relativity (for reviews see \cite{caplip-95,mann}) 
and by the so-called AdS/CFT speculation in string
theory \cite{maldacena}. Very recently, Pretorius and Choptuik 
presented results of fully-nonlinear numerical simulations of the
gravitational collapse of a massless, minimally-coupled scalar field in 2+1
dimensional, axially-symmetric, AdS$_3$ spactime in classical General
Relativity \cite{pretorius-choptuik-00}. Pretorius and Choptuik studied the
critical behavior at the threshold of black-hole formation in this model,
and found a continuous self-similar solution and type-II behavior with a
mass-scaling law with critical exponent $\sim 1.2$. This work was
followed by Garfinkle \cite{garfinkle-00}, who was able to find
analytically an exact solution for this model, in agreement with the
numerical results of Ref.\ \cite{pretorius-choptuik-00}. Husain and
Olivier studied the same model using a different numerical approach and a
different evaluation approach for the critical exponent, and reported on
a mass-scaling law with critical exponent $\sim 1.6$ 
\cite{husain-olivier-00}. This apparent discrepancy of the two
estimates for the critical exponent is of great interest.  

Pretorius and Choptuik also studied in Ref.\ \cite{pretorius-choptuik-00}
the nature of the singularity in their model for supercritical evolutions.
They report on a scalar
curvature singularity, which is spacelike and deformationally strong 
\cite{tipler,ori}. However, the evidence brought in Ref.\ 
\cite{pretorius-choptuik-00} appears to be inconclusive. Specifically,
Pretorius and Choptuik argue that the singualrity is spacelike because the
hypersurface along which the metric variables start growing unboundedly in
the normal direction, and
consequently the curvature invariants begin to diverge, is spacelike.
Although this is consistent with the singularity being spacelike, it is only
a necessary condition for it. The mass inflation
\cite{poisson-israel,ori-91,burko-97,burko-ori} 
null singularity is a
counter-example, where hypersurfaces of constant large (albeit finite)
curvature are spacelike, although the singularity itself is null. The
evidence that
Pretorius and Choptuik bring for the strength of the singularity is that it
is central, namely, that the proper circumference vanishes 
approaching the singularity. This criterion appears indeed to be sufficient
evidence in spherical symmetry under very broad conditions \cite{nolan}, but
this has never been shown in 2+1D and circular symmetry. 

In this paper we show, within a simplified model, that indeed this
singularity is spacelike, scalar curvature, and deformationally strong.
Specifically, we
assume that spacetime asymptotically close to the singularity can be
described by a quasi-homogeneous model, i.e., we assume that spatial
gradients (of the metric functions and the scalar field) are much smaller
than temporal gradients. This assumption seems to be justified by the
results of Ref.\ \cite{pretorius-choptuik-00}, where Figs.\ 18 and 19
imply only mild gradients tangent to the singularity, and much steeper
gradients normal to it. We shall study the
solution to the Einstein-scalar equations in 2+1D and circular symmetry only
asymptotically close to the singularity.   
We thus write the metric as 
\begin{equation}
\,ds^2=N(r)\,dt^2-L(r)\,dr^2+r^2\,d\theta^2, 
\label{metric}
\end{equation}
where $0\le\theta\le 2\pi$ is possibly periodic, and where $N(r)$ and
$L(r)$ are non-negative  
functions (such that $r={\rm const}$ hypersurfaces are spacelike by
construction). The coordinate $r$ is defined such that the
proper circumference of
circles of radius $r$ is $2\pi r$. Note, that $r$ is a timelike coordinate. 
We study the solution to the Einstein-scalar equations 
\begin{equation}
G_{\mu\nu}+\Lambda g_{\mu\nu}=\kappa T_{\mu\nu},
\label{E-scalar}
\end{equation}
where $\Lambda<0$ is the cosmological constant and 
$T_{\mu\nu}=\nabla_{\mu}\phi\nabla_{\nu}\phi
-\frac{1}{2}g_{\mu\nu}\nabla^{\alpha}\phi\nabla_{\alpha}\phi$
is the energy-momentum tensor of the scalar field $\phi$.
Following the convention of Ref.\ \cite{pretorius-choptuik-00} we set the
coupling constant 
$\kappa=4\pi$. The $t-t$ and $r-r$ components of the field equations
(\ref{E-scalar}) are, correspondingly,  
\begin{equation}
L'+2rL^2\Lambda=4\pi rL\phi'^2
\label{e1}
\end{equation}
and
\begin{equation}
N'-2rLN\Lambda=4\pi rN\phi'^2.
\label{e2}
\end{equation}      
These equations are coupled 
to the Klein-Gordon equation for the scalar field, $\Box\phi=0$, 
whose first integral is given by
\begin{equation}
\phi'^2=\frac{L}{N}\frac{d^2}{r^2}.
\label{kg}
\end{equation}
Here, $d^2$ is an integration constant, and a prime denotes differentiation
with respect to $r$. (Recall that we neglect all derivatives with respect to
$t$.) The $\theta-\theta$ component of the field equations is redundant, and
we use it as a consistency check for our solution. This equation is 
\begin{equation} 
LN'^2+NN'L'-2N''NL+4\Lambda N^2L^2=8\pi N^2L\phi'^2.
\label{consistency}
\end{equation} 
We next eliminate $\phi$ from Eqs.\ (\ref{e1}) and (\ref{e2}) using
Eq.\ (\ref{kg}), and find that 
\begin{equation}
L'+2rL^2\Lambda=4\pi d^2 \frac{L^2}{Nr} 
\label{e11}
\end{equation}
and
\begin{equation}
N'-2rLN\Lambda=4\pi d^2 \frac{L}{r}.
\label{e21}
\end{equation}                   
Next, we present a simple solution to Eqs.\ (\ref{e11}) and (\ref{e21}),
which is a generic solution. Then, we show that, in fact, this solution
is the only generic solution. 
Following the analyses of Refs.\ \cite{burko-98,burko-99} we seek a
solution for which 
asymptotically close to the singularity the metric functions behave like 
$L(r)=Ar^{\alpha}$ and $N(r)=Br^{\beta}$, where $A,B>0$ are constants. (At
larger distances from the singularity higher-order terms in $r$ become
important.)  We next seek a
solution to Eqs.\ (\ref{e11}) and (\ref{e21}), and demonstate its validity
by requiring consistency with the fully-nonlinear (and
inhomogeneous) numerical simulations of Ref.\ \cite{pretorius-choptuik-00}. 
Substituting this Ansatz we find first that $\alpha,\beta>0$, and that
asymptotically close to $r=0$ the $\Lambda$-coupled terms are 
negligible. (This was also found in Ref.\ \cite{garfinkle-00}.) Next, we
find
that the solution imposes on $\alpha$ and $\beta$ a relation. Specifically,
we find that $\alpha=\beta$. Also, we find that 
$B=4\pi d^2 A \alpha^{-1}$. Note, that we assume here $\alpha\ne 0$. 
The scalar field satisfied asymptotically close to the singularity
$\phi'^2=\alpha/(4\pi r^2)$. (The case where $\alpha=0$ corresponds to a
vanishing scalar field, and corresponds to the vacuum BTZ solution, for
the case for which $\theta$ is periodic.) 
The dependence of the solution on $\Lambda$ enters only at higher-order
terms $O[r^{2(\alpha+2)}]$. Also, Eq.\ (\ref{consistency}) is 
satisfied by our solution at this order. (When higher-order terms in
$r$ are considered each order will decrease the error associated with the
truncated solution of the lower-order solution.) In fact, one does not
have to assume the Ansatz that $N(r)$ and $L(r)$ are given by simple
powers of $r$. Instead, if one assumes that the $\Lambda$-coupled
terms in Eqs.\ (\ref{e11}) and (\ref{e21}) are negligible (close to
$r=0$), one can solve these equations readily, and obtain that
$L(r)=Ar^{\alpha}$ and $N(r)=4\pi d^2 A \alpha^{-1} r^{\alpha}$ as an
exact and unique solution (with no $\Lambda$ term). Our solution here is
generic in the sense that it relies on the right number of
freely-specifiable parameters (recall that we have the freedom to rescale
the $t$ coordinate, namely $t\to \tilde t=T(t)$, such that $d$ can be set
equal to unity without loss of generality).  

We next show, following Nolan, that this generic solution is the only
generic solution \cite{nolan-com}. 
Defining $X=LN$ and $Y=N/L$, Eqns.\ (\ref{e11}) and (\ref{e21}) 
become
\begin{eqnarray}
X^\prime&=&8\pi\frac{d^2}{r}XY^{-1},\label{eq1}\\
Y^\prime&=&4\Lambda rX^{1/2}Y^{1/2}.\label{eq2}
\end{eqnarray}
Next, we define $u=Y^{1/2}$, and obtain the second order equation
\begin{equation}
ru^{\prime\prime}=(u-4\pi\frac{d^2}{u})^\prime \ ,\label{eq3}
\end{equation}
whose first integral is 
\begin{equation}
ru^\prime = 2u-4\pi\frac{d^2}{u}+4k,\label{eq4}
\end{equation}
where $4k$ is an arbitrary constant. 
Eq.\ (\ref{eq4}) can be separated and written as
$$ \int \frac{u}{u^2+2ku-2\pi d^2}\,du=2\int\frac{dr}{r}. $$
Factorizing the denominator on the left hand side and using undetermined
coefficients gives
$$ \frac{u}{u^2+2ku-2\pi d^2}=
\frac{\beta+k}{2\beta}\frac{1}{u+k+\beta}+\frac{\beta-k}
{2\beta}\frac{1}{u+k-\beta}\ ,$$
which integrates to give logarithmic terms. Then taking the exponential
of both sides gives 
\begin{equation}
|u+k+\beta|^{\frac{\beta+k}{2\beta}}|u+k-\beta|^{\frac{\beta-k}{2\beta}}=e^c
r^2,\label{eq5}
\end{equation}
where $c$ is an arbitrary constant and $\beta^2=k^2+2\pi d^2$. Notice
that $\beta\pm k>0$
(we rule out $d=0$ in which case the scalar field is absent; also $\beta$
means the positive root of $\beta^2$).
Eq.\ (\ref{eq5}) determines $Y(r)$ implicitly and Eqs.\ (\ref{eq2}) and
(\ref{eq4}) can be used to give $X$ in terms of $u$ (or $Y$).
Specifically, we find that 
\begin{equation}
2\Lambda r^2 X^{1/2}=2u-4\pi\frac{d^2}{u}+4k.\label{eq6}
\end{equation}                                                 
 
We determine the asymptotic behavior of $X,Y$ (and hence $N$ and $L$) as
$r\to 0$ as follows. Note that $u\geq 0$. The left hand side of Eq.\ 
(\ref{eq5}) must vanish at $r=0$. Since $u+k+\beta>u+k-\beta$,
we infer that 
$$\lim_{r\to 0}u(r)=\beta-k>0.$$
Thus we can write
\begin{equation}
u(r)=\beta-k+\epsilon(r) \  , \label{eq7}
\end{equation}
where
$$\epsilon(r) = o(1),\qquad r\to 0 \ .$$
Here, $o$ is defined such that  $f(x)=o[g(x)]$ as $x\to 0$ implies
that $f/g\to 0$ as $x\to 0$.   
Substituting this into Eq.\ (\ref{eq5}) yields
\begin{equation}
\epsilon(r)=\epsilon_0 r^{\frac{4\beta}
{\beta -k}}+\epsilon_1(r),\label{eq8}
\end{equation}
where $\epsilon_1=o(\epsilon)$ as $r\to 0$ and $\epsilon_0$ is a
constant which is
determined by $\beta, k$ and $c$. Equations (\ref{eq7}) and (\ref{eq8})
are sufficient to determine
the leading order behavior of $X$ via Eq.\ (\ref{eq6}), i.e., 
$$X^{1/2}=\frac{2\beta}{\Lambda(\beta-k)}\epsilon_0
r^{2(\frac{\beta+k}{\beta-k})} + o(\epsilon) \ .$$
Then the original metric functions $N,L$ satisfy
\begin{eqnarray}
N&=&uX^{1/2}\sim\frac{2\beta}{\Lambda}
\epsilon_0r^{2(\frac{\beta+k}{\beta-k})},
\qquad r\to0,\label{eq9}\\
L&=&\frac{X^{1/2}}{u}\sim\frac{2\beta}{\Lambda(\beta-k)^2}
\epsilon_0r^{2(\frac{\beta+k}{\beta-k})},\qquad
r\to 0.
\label{eq10}
\end{eqnarray}                                            
This proves that in general, the metric functions $N,L$ of 
these space-times display power-law behavior as the singularity is
approached, with the same power of $r$ in each function. [The 
$\Lambda^{-1}$ in Eq.\ (\ref{eq9}) and Eq.\ (\ref{eq10}) 
can be absorbed into $\epsilon_0$ by re-scaling the $t$ coordinate.] 
Also, we find that $\alpha=2(\beta+k)/(\beta-k)$. 

With this solution we first show that the singularity is scalar polynomial.
Note that all the following expression are given to leading order in $r$. 
Specifically, we find the Kretschmann scalar 
$R_{\mu\nu\rho\sigma}
R^{\mu\nu\rho\sigma}=3\alpha^2/(A^2r^{2\alpha+4})$, and the Ricci scalar 
$R=\alpha/(Ar^{\alpha+2})$. Both curvature scalars diverge approaching $r=0$. 
Next, we show that this singularity is strong
in the sense of Tipler \cite{tipler} (or deformationally strong in the sense
of Ori \cite{ori}). 
The timelike-timelike component of the Ricci tensor in a parallel-propagated
frame is given by $R_{(0)(0)}=\alpha/(Ar^{\alpha+2})$. Re-expressing that in
terms of the proper time of a radial observer (who follows a $t={\rm const}$
timelike geodesic), we find that 
\begin{equation}
R_{(0)(0)}(\tau)=4\frac{\alpha}{(\alpha+2)^2}\frac{1}{\tau^2},
\label{ricci}
\end{equation}  
where $\tau$ is (future directed) proper time, set such that $\tau\to 0^-$
approaching the singularity. Here, and in what follows, we find 
$R_{(0)(0)}(\tau)$ to leading order in $\tau^{-1}$. [The geodesic
equation for this geodesic is $\ddot{r}+(\alpha/2r)\dot{r}^2=0$, whose
solution is $\tau(r)=-2A^{1/2}(\alpha+2)^{-1}r^{(\alpha+2)/2}$, where an
overdot denots differentiation with respect to $\tau$.]   
Note that $R_{(0)(0)}(\tau)$ is independent of
$A$. [Also $R_{\mu\nu\rho\sigma}
R^{\mu\nu\rho\sigma}(\tau)=48\alpha^2(\alpha+2)^{-4}\tau^{-4}$ is
independent of $A$.] This is similar to the independence of
$R_{(0)(0)}(\tau)$ of the density (or pressure) in the
Friedmann-Robertson-Walker cosmology near the singularity, or the
independence of the Kretschmann scalar of the mass in Schwarzschild, or of
the density in the Friedmann-Robertson-Walker cosmology \cite{burko-book}. 

We next use a theorem by Clarke and Kr\'{o}lak, according to which a
sufficient condition for the singularity to be strong in the sense of Tipler
is that 
\begin{equation}
\int^{\tau}\,d\tau '\int^{\tau '}\,d\tau '' \, R_{(0)(0)}(\tau '')
\end{equation}
diverges as $\tau\to 0^-$ \cite{clarke-krolak}. From Eq.\ (\ref{ricci}) it
is clear that this is indeed the case, as this quantity diverges
logarithmically in $\tau$ as $\tau\to 0^-$,  
such that we show that the singularity is indeed
deformationally strong. (Note that the proof of the Clarke-Kr\'{o}lak
theorem can be easily extended to 2+1 dimensions.) 

Finally, we make the following remarks on the applicability of our 
homogeneous model. First, we find that asymptotically close to the
singualrity the scalar field diverges like 
\begin{equation}
\phi(r)=\left(\frac{\alpha}{4\pi}\right)^{1/2}\ln r.
\end{equation}
Indeed, the fully-nonlinear and inhomogeneous simulations find this
logarithmic divergence for $\phi(r,t)$ \cite{pretorius-communication}.
This, however, provides us also with a quantitative check for the
predictions of the homogeneous model. Specifically, the amplitude 
$(\alpha/4\pi)^{1/2}$ of the scalar field involves the same parameter 
$\alpha$ as in the metric functions $N,L$. Consequently, our homogeneous
model
captures the pointwise behavior at the singularity well if fully-nonlinear
and inhomogeneous simulations confirm our prediction that 
\begin{equation}
\left(\frac{r\,\partial L(r,t)/\,\partial r}{4\pi L(r,t)}\right)^{-1/2}
\phi(r,t)\frac{1}{\ln r}\longrightarrow 1
\end{equation}
as $r\to 0$ towards any point along the singularity. Second, the  numerical
simulations also indicate that the Ricci scalar grows approaching the
singularity like $R(\tau)\approx \tau^{-2}$ \cite{pretorius-communication}.
Indeed, in our model we find that 
$R(\tau)=4\alpha(\alpha+2)^{-2}\tau^{-2}$. In addition to the right behavior
as a function of proper time, we also predict a specific dependence on 
$\alpha$, which can be checked numerically. 
Finally, the quasi-homogeneity
we assume is supported by the simulations in Ref.\
\cite{pretorius-choptuik-00}, where only mild (possibly
oscillatory) dependence on $t$ is reported. 

We showed that in a quasi-homogeneous model for the singularity in 2+1D AdS
black hole with a self-gravitating scalar field in circular symmetry, the
singularity is spacelike, scalar curvature, and deformationally strong. We
believe that this simple model captures the pointwise behavior near the
singularity when one allows for inhomogeneities. The question of how the
solution is modified by inclusion of inhogeneities or angular momentum
remains open.  

I am indebted to Matt Choptuik, Brien Nolan, and Frans Pretorius for
useful discussions. This research was supported by NSF grant No.\
AST-9731698.

\end{document}